\documentclass[
    onecolumn,
    10pt,
    a4paper,
    float,
    aps,
    pra,
    nofootinbib
]{revtex4-2}
\makeatletter
\def\pdfstartlink@attr{}
\makeatother

\usepackage{amsmath}
\usepackage{amssymb}
\usepackage{amsthm}

\usepackage[utf8]{inputenc}

\usepackage[dvipsnames]{xcolor}
\usepackage{hyperref}
\hypersetup{
    colorlinks  =   true,
    citecolor   =   BrickRed,
    linkcolor   =   RoyalBlue,
    urlcolor    =   RoyalBlue
    }
\DeclareUrlCommand\doi{\def\UrlLeft##1\UrlRight{\href{https://dx.doi.org/##1}{##1}}}

\usepackage{float}
\usepackage{graphicx}
\usepackage{ulem}

\usepackage{soul}

\usepackage{bm}
\usepackage{bbm}
\usepackage{bbold}
\usepackage{multirow}
\DeclareMathAlphabet{\mathbbb}{U}{bbold}{m}{n}

\usepackage{mathrsfs}

\newcommand{\tp}{\mathsf{T}}

\DeclareMathOperator{\haf}{haf}
\DeclareMathOperator{\lhaf}{lhaf}

\begin{document}

\title{\large Derivation of the Loop Hafnian Generating Function \\ for Arbitrary Symmetric Matrices via Gaussian Integration}

\author{S.~V.~Tarasov
}
\email{serge.tar@gmail.com}
\noaffiliation

\date{\today}

	

\begin{abstract}
This short note shows that the recently proposed generating function for loop hafnians---originally derived using quantum-optical methods for a restricted class of matrices---is in fact valid for arbitrary symmetric matrices. 
The proof relies solely on Gaussian integration and does not assume any additional properties inherited from the covariance matrices of quantum Gaussian states.
\end{abstract}

\maketitle

Hafnians, denoted by $\haf S$, and loop hafnians, denoted by $\lhaf S$, are closely related polynomial functions of entries of a symmetric matrix $S$.
They arise in multivariate calculations involving the Gaussian measure due to the combinatorial reasons.
Thus, they naturally appear in the fields of probability theory, statistical physics of many-body systems, quantum field theory, and quantum optics.
The exact---and even approximate, under sufficient restrictions---calculation of hafnians and loop hafnians is believed to be a problem in the $\sharp P$-hard complexity class for a general case matrix.
Because of this, the hafnians and loop hafnians play an important role in studies of quantum computations and simulations.

The recent preprint \cite{lHMT-arxiv24} revealed the generating function for the loop hafnians.
It was derived via the methods of quantum optics and thus is guaranteed to be applicable only for matrices corresponding to some valid  ``quantum Gaussian states'' (which means, there should be some special block symmetries).
Throughout the analysis of~\cite{lHMT-arxiv24}, which focuses on the capabilities of simulating quantum optical circuits, the found generating function is heavily employed. 
In several instances, the relevant matrices fall outside the class for which the formula for the generating function was rigorously justified.
In such cases, in order to take advantage of this formula, the authors treat its applicability as an assumption, supported by numerical tests.

The purpose of this short note is to share a simple proof ensuring that the aforementioned loop hafnian generating function is valid for an {\it arbitrary} symmetric matrix.
The proposed proof is based on the standard Gaussian integration, which underlies the quantum optics techniques when operating with quantum Gaussian states.
The note consists of several relevant definitions, the statement of the theorem, a well-known lemma useful for the loop hafnian calculation, and the announced proof itself.

\vspace{1.0\baselineskip}
\noindent {\bf Definition 0.} 
In what follows, the symbols $\mathbb{1}$ and $\mathbb{0}$ denote the identity and zero matrices, respectively. 
The matrix $J_{n\times m}$ denotes an $n\times m$ matrix with all entries equal to $1$. 
Bold symbols represent vectors.
The superscript $(\ldots)^\tp$ denotes matrix or vector transposition.

\vspace{1.0\baselineskip}
\noindent {\bf Definition 1.} A {\it hafnian} of a $2m\times 2m$ symmetric matrix $S = (s_{i,j})$ is the following polynomial in its entries:
\begin{equation}    \label{def-haf}
    \haf S  =   \sum_{\pi \in \text{PMP}(2m)} \prod_{\{i,j\}\in \pi} s_{i,j}
    \ = \ \sum_{\sigma} \, \prod_{j=1}^m s_{\sigma_{2j-1},\sigma_{2j}}.        
\end{equation}
Here the sum over partitions $\pi$ runs over a set of perfect matching permutations $\text{PMP}(2m)$, which collects all ways to partition the set $\{1,\ldots,2m\}$ into unordered pairs $\{i,j\}$.
Equivalently, the sum can be expressed over permutations~$\sigma$ of the set $\{1,\ldots,2m\}$, such that $\sigma_{2j-1}<\sigma_{2j}$ and $\sigma_{2j-1}<\sigma_{2j+1}$ for all $j$.

\vspace{1.0\baselineskip}
\noindent {\bf Definition 2.1.} A {\it loop hafnian} of a $\mu\times\mu$ symmetric matrix $S = (s_{i,j})$ is another polynomial in the entries of $S$: 
\begin{equation}    \label{def-lhaf-pre}
    \lhaf S
    = \sum_{\pi \in \text{SPM}(\mu)} 
        \prod_{\substack{B\in \pi:\\ B=\{i,j\}}} \!\! s_{i,j} \
        \prod_{\substack{B\in \pi:\\ B=\{i\}}} \!\! s_{i,i}.
\end{equation}
The partition $\pi$ runs over a set of single-pair matchings $\text{SPM}(\mu)$, which collects all ways to partition the set $\{1,\ldots, \mu\}$ into pairs or singletons;
the blocks $B$ of the partition $\pi$ may be either $\{i,j\}$ or $\{i\}$.
Unlike the hafnian, Eq.~(\ref{def-haf}), the loop hafnian depends on the diagonal entries of the matrix, and is a well-defined function for both even- and odd-dimensional matrices.
For the matrices with zero diagonal entries, these two matrix functions coincide. 

The term ``loop hafnian'' was originally introduced in \cite{Bjorklund-2019-lhaf0}, 
which also provides transparent graph-theory illustrations of the~$\text{SPM}$ set (fig.~1 therein).
In quantum mechanics, loop hafnians appear as matrix elements of general multimode Gaussian state density matrices in the Fock basis \cite{Quesada-PRA2019-lhaf}.

\vspace{1.0\baselineskip}
\noindent {\bf Definition 2.2} (as generalized in \cite{lHMT-arxiv24}). 
A {\it loop hafnian} of a $\mu\times\mu$ symmetric matrix $S = (s_{i,j})$ and $\mu$-length vector ${\bf v} = (v_1,v_2,\ldots, v_\mu)$ is the following polynomial in entries of $S$ and ${\bf v}$:
\begin{equation}    \label{def-lhaf}
    \lhaf \big(S, {\bf v}\big) = \sum_{\pi \in \text{SPM}(\mu)} 
        \prod_{\substack{B\in \pi:\\ B=\{i,j\}}} \!\! s_{i,j} \
        \prod_{\substack{B\in \pi:\\ B=\{i\}}} \!\! v_i . 
\end{equation}
Unlike the previous definition, Eq.~(\ref{def-lhaf-pre}), a singleton $\{i\}$ contributes $v_i$ rather than $s_{i,i}$.
The vector ${\bf v}$ is called ``the loop vector''.
Two definitions coincide if the entries of ${\bf v}$ match the diagonal entries of $S$, that is, $S_{i,i} = v_i$ for  $i=1,\ldots,\mu$. 
For a zero ``loop vector'' {\bf v}, the loop hafnian $\lhaf \big(S, {\bf v}\big)$ is reduced to the hafnian $\haf S$.

\vspace{1.0\baselineskip}
\noindent {\bf Theorem.} {\it The following loop hafnian master theorem provides the generating function for the loop hafnians and holds for any symmetric $2m \times 2m$ complex matrix $S = S^\tp$ complemented by any $2m$-dimensional complex vector ${\bf v}$:}
\begin{equation}    \label{lHaf-theorem}
    \frac{
        \exp\Big(\frac{1}{2} {\bf v}^\tp \big( \mathbb{1}-\mathbb{Z}\,S \big)^{-1} \, \mathbb{Z} \, {\bf v} \Big)
    }{
        \sqrt{\,\det\big(\mathbb{1} - \mathbb{Z}\,S\big)\,}
    }
    = 
    \sum_{\{n_j\}} \lhaf \Big(\tilde{S}_{\{n_j\},\{n_j\}};\tilde{\bf v}_{\{n_j\},\{n_j\}} \Big) 
    \prod_{j=1}^m \frac{z_j^{n_j}}{n_j!};
    \qquad
    \mathbb{Z}  \equiv  \begin{bmatrix}   \mathbb{0}  &   \text{diag}\big(\{z_j\}\big) \\
                                \text{diag}\big(\{z_j\}\big)    &   \mathbb{0}  \end{bmatrix}.
\end{equation}
{\it
\noindent
The sum on the right-hand side runs over all $m$-tuples of non-negative integers $\{n_j\}$.
The $2m\times 2m$ matrix $\mathbb{Z}$ contains a pack of $m$ variables $\{z_j\}$---complex, in general case!---as diagonal entries of the counter-diagonal blocks, so each argument~$z_j$
appears twice every time the matrix $\mathbb{Z}$ enters the function on the left-hand side.
The matrix $\tilde{S}_{\{n_j\},\{n_j\}}$ of the $\big(2\sum_{j=1}^m n_j\big) \times \big(2\sum_{j=1}^m n_j\big)$ dimension is constructed from the matrix $S$ as follows: each element $s_{i,j}$, $1\le i,j\le m$, is replaced by an $n_i\times n_j$ block $s_{i,j} \, J_{n_i\times n_j}$ filled with copies of $s_{i,j}$; the similar replacement is done for the elements $s_{i,j+m}$, $s_{i+m,j}$, $s_{i+m,j+m}$ as well.
The~$\big(2\sum_{j=1}^m n_j\big)$-dimensional vector $\tilde{\bf v}_{\{n_j\},\{n_j\}}$ is constructed from the vector ${\bf v}$ whose entries $v_j$ and $v_{j+m}$ are repeated $n_j$~times, respectively.
By convention, the loop hafnian on the right-hand side is unity for trivial $m$-tuples $\{n_j = 0\}$, $j=1,\ldots, m$.
This convention also determines the branch of the square root function that yields the value $\sqrt{\det(\mathbb{1} - \mathbb{Z} S)}\big|_{\{z_j=0\}} = +1$ at the origin of the complex variables space $\{z_j\}$.}
\\

The generating function for the loop hafnians was derived in the preprint \cite{lHMT-arxiv24} within a quantum-optical framework.
The key idea of the derivation was based on the fact that the probabilities of measuring the sets of photon numbers when sampling from a quantum Gaussian state with nonzero displacement are proportional to the loop hafnians of certain matrices (see Eq.~(22) in \cite{Hamilton2019-DetailedGBS}) and the sum of these probabilities over all possible outcomes is unity.
The outcome probabilities are expressed in terms of the loop hafnians via a proper use of the Husimi $Q$-representation of the quantum state and the Glauber–Sudarshan $P$-representations of the projection measurement operators.
In~this formulation, matrices and vectors that are arguments of the loop hafnians are always related to the covariance matrices of some quantum Gaussian state, which implies several special symmetries of their blocks.
However, the quantum specificity introduced by the employed framework is not fundamental, since all implemented operations for the considered quantum states are, at their core, manifestations of Gaussian integration.

\vspace{1.0\baselineskip}
\noindent{\bf Lemma\,\cite{Quesada-PRA2019-lhaf}.}
{\it 
Let $S$ be a symmetric  $\mu \times \mu$ matrix, and let ${\bf v}$ be a $\mu$-dimensional vector.
Consider a sequence of $\mu$~non-negative integers, $\{n_1,\ldots, n_\mu\}$.
Define an extended matrix $\tilde{S}_{\{n_j\}}$ as the $\big(\sum_{j=1}^\mu n_j\big)\times\big(\sum_{j=1}^\mu n_j\big)$ matrix, such that each entry $s_{i,j}$ of the matrix $S$ is replaced by an $n_i\times n_j$ block $s_{i,j}\,J_{n_i\times n_j}$ filled with $s_{i,j}$.
Similarly, define an extended vector $\tilde{\bf{v}}_{\{n_j\}}$ of the length $\big(\sum_{j=1}^\mu n_j\big)$, which is obtained by repeating each entry $v_j$ of the vector ${\bf v}$ exactly $n_j$ times. 
Introduce also a vector ${\bf x} = (x_1,\ldots,x_\mu)$ of $\mu$ variables $x_j$.
The loop hafnian may be calculated as follows:
}
\begin{equation}    \label{lhaf-lemma}
    \lhaf \big(\tilde{S}_{\{n_j\}};\tilde{\bf v}_{\{n_j\}}\big) 
    = \left[ \prod_{j=1}^{\mu} \frac{\partial^{n_j} \ }{\partial        x_j^{n_j}} \right] \exp\left(\frac{1}{2}{\bf x}^\tp S \, {\bf x}   + {\bf v}^\tp {\bf x}\right) \Bigg|_{\{x_j=0\}} \!\!
    = \left[ \, \prod_{r=1}^{\sum_{j=1}^\mu n_j} 
        \frac{\partial \ \ }{\partial x_{\alpha_r}} \, \right]
        \exp\left(\frac{1}{2}{\bf x}^\tp S \, {\bf x} + {\bf v}^\tp {\bf x} \right) 
    \Bigg|_{\{x_j=0\}} \!
    .
\end{equation}
(The second, alternative form has only first-order partial derivatives $\partial / \partial x_j$ in the product, and each of them is applied $n_j$ times in a row.
The product here runs over the index $r$, 
with $\{\alpha_r\}$ being a non-decreasing sequence where each entry $j=1,\ldots,\mu$ appears $n_j$ times.)

\vspace{0.5\baselineskip}
Note that the extended matrix $\tilde{S}_{\{n_j\},\{n_j\}}$ and vector $\tilde{\bf v}_{\{n_j\},\{n_j\}}$ entering the theorem, Eq.~(\ref{lHaf-theorem}), are special cases of the extended matrix $\tilde{S}_{\{n_j\}}$ and vector $\tilde{\bf v}_{\{n_j\}}$ introduced in the lemma.

The statement of the lemma is proved in the appendix A of \cite{Quesada-PRA2019-lhaf}.
An analogous result for the hafnians (corresponding to a zero ``loop vector'' case) has been previously derived in \cite{Hamilton2017-GBS,Hamilton2019-DetailedGBS} on the basis of similar ideas.
For the sake of completeness, a comment following their arguments is provided below.
\\

Consider the right-hand side and take the mixed derivative according to the multivariate Fa\`a di Bruno's formula~\cite{Hardy2006FDB}:
\begin{equation}    \label{Faa-di-Bruno}
    \left[ \, \prod_{r=1}^{\sum_{j=1}^\mu n_j} 
        \frac{\partial \ \ \, }{\partial x_{\alpha_r}} \, \right]
    \exp\big(p({\bf x})\big)
    =
    \sum_\pi \frac{\partial^{|\pi|} \exp(p)}{\partial p^{|\pi|}} \prod_{B\in\pi} 
            \frac{\partial^{|B|}\,p({\bf x})}{\prod_{r\in B}\partial x_{\alpha_r} };
    \qquad
    p({\bf x}) \equiv \frac{1}{2}{\bf x}^\tp S \,{\bf x} + {\bf v}^\tp {\bf x}.
\end{equation}
Here $\pi$ runs over all possible partitions of the set $\{1,2,3,\ldots, \sum_{j=1}^\mu n_j\}$,
$B$ denote blocks of the partition $\pi$, and $|\ldots|$ means the cardinality of the set.
Since the ``outer'' function is exponential, all derivatives with respect to  $p$ yield the same common factor, which equals 1 when evaluated at the point $\{x_j=0\}$.
The ``inner'' function $p({\bf x})$ is quadratic in $x_j$, thus only partitions $\pi$ in blocks of size 1 and 2---singletons and pairs---produce nonzero contributions:
\begin{equation}
     B = \{i\}, \ |B| = 1\!: \quad  \partial\, p({\bf x})\, \big/ \, \partial\,x_{\alpha_i} = {\bf v}_{\alpha_i} + \big( S \, {\bf x} \big)_{\alpha_i};
    \qquad \ \
    B = \{i,k\}, \ |B| = 2\!: \quad \ \partial^2\, p({\bf x})\, \big/ \, \partial\,x_{\alpha_i} \, \partial\,x_{\alpha_k} = s_{\alpha_i,\alpha_k}.
\end{equation}
Thus, the sum in Eq.~(\ref{Faa-di-Bruno}), being evaluated in the origin $\{x_j = 0\}$, is exactly the same as the one that defines the loop Hafnian of $\tilde{S}_{\{n_j\}}$ and $\tilde{\bf v}_{\{n_j\}}$  according to Eq.~(\ref{def-lhaf}). 
\\

\noindent{\bf Proof of the theorem.}
Technically, the proof closely follows that of the hafnian master theorem \cite{LAA2022-HafnianMasterTheorem}.
The only difference is that the hafnian master theorem deals with Gaussian integrals whose exponent is free of linear term, while generalization to loop hafnians introduces such linear shift explicitly.

A convenient starting point is given by the lemma formulated above, see Eq.~(\ref{lhaf-lemma}):
\begin{equation}    \label{proof-start}
    \lhaf \big(\tilde{S}_{\{n_j\},\{n_j\}};\tilde{\bf v}_{\{n_j\},\{n_j\}}\big) 
        = \Bigg[
            \prod_{j=1}^m \bigg( \frac{\partial \ }{\partial x_j} \frac{\partial \ }{\partial x_{j+m}}\bigg) ^{n_j}
        \Bigg] \,
        g({\bf x}) \Bigg|_{\{x_j=0\}}\!,
        \qquad
        g({\bf x}) \equiv \exp\left(\frac{1}{2}{\bf x}^\tp S {\bf x} + {\bf v}^\tp {\bf x}\right).
\end{equation}
\\

Assume the symmetric matrix $S\in\mathbb{C}^{2m\times 2m}$ is diagonalizable and invertible.
Let the matrix $S^{1/2}$ be any of its square roots, which is also a diagonalizable and invertible $2m\times 2m$ complex symmetric matrix.
In order to employ the Gaussian-integral technique, let us represent the function $g({\bf x})$ as follows:
\begin{equation}    \label{g-Gauss-int}
    g({\bf x}) = \frac{\exp\left(-\frac{1}{2}{\bf v}^\tp S^{-1} {\bf v}\right)}{(2\pi)^m}
        \int_{\mathbb{R}^{2m}} 
            \exp\left(-\frac{1}{2} \boldsymbol\xi^\tp \boldsymbol\xi + {\bf v}^\tp S^{-1/2} \boldsymbol\xi + {\bf x}^\tp S^{1/2} \boldsymbol{\xi}\right) \, d^{2m} \boldsymbol\xi.
\end{equation}
This is a standard identity following from the Gaussian integral with a shifted mean. 
The differential operator in Eq.~(\ref{proof-start}), being applied to the representation (\ref{g-Gauss-int}) of $g({\bf x})$, only acts on the term  $\exp\left({\bf x}^\tp S^{1/2} \boldsymbol{\xi}\right)$ and generates the polynomial prefactor in the integrand.
However, this prefactor may be generated in an alternative way, via an auxiliary Gaussian exponent:
\begin{equation}    \label{hint}
\begin{split}
    &\Bigg[
        \prod_{j=1}^m \bigg( \frac{\partial \ }{\partial x_j} \frac{\partial \ }{\partial x_{j+m}}\bigg) ^{n_j}
    \Bigg]\exp\left({\bf x}^\tp S^{1/2} \boldsymbol{\xi}\right) \Bigg|_{\{x_j=0\}}
    =
    \prod_{j=1}^m \left((S^{1/2}\boldsymbol\xi)_j (S^{1/2}\boldsymbol\xi)_{j+m} \right)^{n_j}
    \\
    &\qquad\qquad=
    \Bigg[\prod_{j=1}^m \frac{\partial^{n_j} }{\partial z_j^{n_j}}\Bigg] 
        \exp \left(
            \frac{1}{2} (S^{1/2}\boldsymbol\xi)^\tp \mathbb{Z} (S^{1/2}\boldsymbol\xi)
                \right)\Bigg|_{\{z_j=0\}}\!,
    \qquad
    \mathbb{Z} \equiv \begin{bmatrix}   \mathbb{0}  &   \text{diag}\big(\{z_j\}\big) \\
                                \text{diag}\big(\{z_j\}\big)    &   \mathbb{0}  \end{bmatrix}.
\end{split}
\end{equation}
Here appears the block-counter-diagonal matrix $\mathbb{Z}$ of dimension $2m\times 2m$, whose elements $\{z_j\}$ are in fact arguments of a multivariate generating function to be derived.
That standard hint is a central element of the whole proof.
It becomes possible since the differential operator on the left-hand side of Eq.~(\ref{hint}) is symmetric with respect to each pair of variables $\{x_j,x_{j+m}\}$ and the corresponding derivatives are always applied in the same order.
Updating the expression in Eq.~(\ref{proof-start}) for the loop hafnian according to Eqs.~(\ref{g-Gauss-int})--(\ref{hint}) yields:
\begin{equation}    \label{p-stage2}
\begin{split}
    \lhaf \big(\tilde{S}_{\{n_j\},\{n_j\}};\tilde{\bf v}_{\{n_j\},\{n_j\}}\big) 
        \! = \!
        &\Bigg[\prod_{j=1}^m \frac{\partial^{n_j} }{\partial z_j^{n_j}}\!\Bigg] \!
                \frac{\exp\left(-\frac{1}{2}{\bf v}^\tp S^{-1} {\bf v}\right)}{(2\pi)^m}
        \!\int_{\mathbb{R}^{2m}}\!\!\!\!
        \exp\!\left(\!\!-\frac{1}{2}\boldsymbol\xi^\tp \big(\mathbb{1} - S^{1/2} \mathbb{Z} S^{1/2}\big) \boldsymbol\xi + {\bf v}^\tp \! S^{-1/2}\boldsymbol\xi \!\right) \!d^{2m}\boldsymbol\xi
        \, \Bigg|_{\{z_j=0\}}\!\!.
\end{split}        
\end{equation}
The differential operator should act at the origin of the complex variables space $\{z_j\}$.
Thus, it is sufficient to consider the relation~(\ref{p-stage2}) only in the vicinity of the point $\{z_j=0\}$.

So, let the absolute values of the arguments $\{z_j\}$ be small enough to ensure that the real part of the matrix $\big(\mathbb{1} - S^{1/2} \,\mathbb{Z}\, S^{1/2}\big)$ in the integrand's exponent is positive definite.  
Then, the Gaussian integral over $\{\xi_j\}$ is calculated in a straightforward way:
\begin{equation}
\begin{split}
    \lhaf \big(\tilde{S}_{\{n_j\},\{n_j\}};\tilde{\bf v}_{\{n_j\},\{n_j\}}\big)
        =
        &\Bigg[\prod_{j=1}^m \frac{\partial^{n_j} }{\partial z_j^{n_j}}\Bigg] \,
                \frac{\exp\Big(-\frac{1}{2} {\bf v}^\tp \big( S^{-1} - S^{-1} (S^{-1} -\mathbb{Z})^{-1}      
                    S^{-1}\big) {\bf v} \Big)
                    }
                    {
                        \sqrt{\det\big(\mathbb{1} - S^{1/2}\,\mathbb{Z}\,S^{1/2}\big)\,}}\, \Bigg|_{\{z_j=0\}
                    }.
\end{split}        
\end{equation}
It remains to simplify the expression by reorganizing the matrix product in the determinant and eliminating the inverse matrices $S^{-1}$ in the numerator.
The latter could be done via elementary matrix algebra:
\begin{equation}
    S^{-1} - S^{-1} (S^{-1}-\mathbb{Z})^{-1}S^{-1} 
    = \big(\mathbb{1} - (\mathbb{1}-\mathbb{Z} S)^{-1} \big) S^{-1} 
    = -(\mathbb{1}-\mathbb{Z}S)^{-1} \mathbb{Z} S\, S^{-1} 
    = -(\mathbb{1}-\mathbb{Z}S)^{-1} \mathbb{Z}.
\end{equation}
The final result is:
\begin{equation}
\begin{split}
    \lhaf \big(\tilde{S}_{\{n_j\},\{n_j\}};\tilde{\bf v}_{\{n_j\},\{n_j\}}\big)
        =
        &\Bigg[\prod_{j=1}^m \frac{\partial^{n_j} \ }{\partial z_j^{n_j}}\Bigg] \,
            \frac{
                \exp\Big(\frac{1}{2} {\bf v}^\tp \big( \mathbb{1}-\mathbb{Z}\,S \big)^{-1}\, \mathbb{Z} \, {\bf v} \Big)
            }{
                \sqrt{\det\big(\mathbb{1} - \mathbb{Z}\,S\big)\,}
            }\, 
            \Bigg|_{\{z_j=0\}}.
\end{split}        
\end{equation}
The loop hafnians $\lhaf \big(\tilde{S}_{\{n_j\},\{n_j\}};\tilde{\bf v}_{\{n_j\},\{n_j\}}\big)$ are indeed partial derivatives of the function on the left-hand side of the Eq.~(\ref{lHaf-theorem}).

Although the proof assumes that the matrix $S$ is diagonalizable and invertible, this is not essential.  
Note that both sides of the Eq.~(\ref{lHaf-theorem}) are continuous functions in matrix entries $s_{i,j}$, provided that the variables $\{z_j\}$ lie in a sufficiently small neighborhood of the origin $\{z_j=0\}$.
Since any symmetric matrix $S$ can be approximated arbitrarily closely by a diagonalizable and invertible one, the result extends to all symmetric even-size matrices.
\qedsymbol{}
\\

\noindent
{\bf Remark 1.} The matrix hafnians, Eq.~(\ref{def-haf}), are perhaps best known for their relation to Wick's (or Isserlis's) theorem~\cite{Isserlis1918,Wick1950}.
They encode {\it the joint central moments} of a set of stochastic variables $\{X_1,\ldots,X_{2m}\}$, distributed according to a Gaussian measure with a covariance matrix~$K_{XX}$: 
\begin{equation}
    \mathbb{E}\big(\delta X_1 \cdot \delta X_2 \cdot \ldots \cdot \delta X_{2m}\big) = \haf K_{XX}, \quad \text{where} \quad \delta X_j \equiv X_j - \mathbb{E}(X_j).
\end{equation}
The generalized definition 2.2 of the loop hafnians, Eq.~(\ref{def-lhaf}), enables similar intuition regarding the probability theory.
Namely, the loop hafnians are representations of {\it the joint initial (non-centered) moments} of stochastic variables $\{X_1,\ldots,X_\mu\}$, distributed according to a Gaussian measure with a covariance matrix~$K_{XX}$: 
\begin{equation}    \label{lhaf-Wick}
    \mathbb{E}\big(X_1 \cdot X_2 \cdot \ldots \cdot X_{\mu}\big) = \lhaf \big(K_{XX}, \mathbb{E}({\bf X})\big),
    \quad \text{where} \quad 
    \mathbb{E}\big({\bf X}\big) \equiv \big(\mathbb{E}(X_1), \mathbb{E}(X_2), \ldots, \mathbb{E}(X_\mu)\big).
\end{equation}

Indeed, assume the $\mu\times\mu$ symmetric matrix $S$ is real and positive definite and thus may be interpreted as a valid covariance matrix.
Identifying ${\bf X} = S^{1/2} \boldsymbol\xi$ allows the statement of lemma, Eq.~(\ref{lhaf-lemma}), combined with the Gaussian integral representation of $g$ given in Eq.~(\ref{g-Gauss-int}) (with $2m$ being replaced by an arbitrary integer $\mu$), to be immediately rewritten in the form
\begin{equation}
    \lhaf \big(S, {\bf v}  \big) = \int_{\mathbb{R}^\mu} \bigg(\prod_{j=1}^{\mu}X_j \bigg) \, \rho \big({\bf X} \big)\, d^{\mu}{\bf X},
    \qquad
    \rho \big({\bf X} \big) \equiv \frac{\exp \big( - \frac{1}{2} ({\bf X}-{\bf v})^\tp S^{-1} ({\bf X}-{\bf v}) \big)}{(2\pi)^{\mu/2} \sqrt{\det S}},
    \quad
    {\bf v} \equiv \mathbb{E}({\bf X}),
\end{equation}
where the right-hand side is literally a definition of joint moments with respect to the probability density $\rho({\bf X})$.

The Eq.~(\ref{lhaf-Wick}) would serve a nice alternative starting point for the proof, which allows one to shorten it even more by starting from Eq.~(\ref{p-stage2}).
The generalization to any symmetric matrix $S$ would be straightforward since both sides of the generating function identity are analytical functions on $s_{i,j}$ and $v_j$ (provided the arguments $\{z_j\}$ are small enough).
\\


\noindent
{\bf Remark 2.}
The generating function in Eq.~(\ref{lHaf-theorem})  may appear somewhat restrictive and non-universal since the {\it even dimension} of the matrix $S$ is explicitly pronounced in the generating function structure, while the loop-hafnian function is well defined for square symmetric matrices of {\it any size}.
Thus, all the loop-hafnians of odd-dimensional matrices seem to be excluded from the consideration.
One way to correct this unfair situation is to employ the following relation between loop hafnians of even-sized and odd-sized matrices:
\begin{equation}
    \lhaf S^{(2n-1)\times(2n-1)}
    =
    \lhaf 
        \left[
        \begin{array}{c|ccc} 
        1 & 0 & \cdots & 0 \\
        \hline
        0 & & &  \\
        \vdots & & S^{(2n-1)\times(2n-1)} & \\
        0 & & & \\
        \end{array}
        \right].                                          
\end{equation}
Now, the loop hafnians of even-dimensional matrices are described by the generating function as well.
\\

\noindent
{\bf Remark 3.} 
It is interesting to note that the generating function appearing in the hafnian master theorem~\cite{LAA2022-HafnianMasterTheorem} was originally derived using quantum-optical methods as well. 
It was derived while studying the statistical distribution of atoms in a weakly interacting Bose-Einstein-condensed gas over multiple non-condensate modes~\cite{PRA2022-AtomicBS}. 
Naturally, the generating function in \cite{PRA2022-AtomicBS} had been restricted to the hafnians of the covariance-related matrices only.

\bibliography{list}

\begin{thebibliography}{10}%
\makeatletter
\providecommand \@ifxundefined [1]{%
 \@ifx{#1\undefined}
}%
\providecommand \@ifnum [1]{%
 \ifnum #1\expandafter \@firstoftwo
 \else \expandafter \@secondoftwo
 \fi
}%
\providecommand \@ifx [1]{%
 \ifx #1\expandafter \@firstoftwo
 \else \expandafter \@secondoftwo
 \fi
}%
\providecommand \natexlab [1]{#1}%
\providecommand \enquote  [1]{``#1''}%
\providecommand \bibnamefont  [1]{#1}%
\providecommand \bibfnamefont [1]{#1}%
\providecommand \citenamefont [1]{#1}%
\providecommand \href@noop [0]{\@secondoftwo}%
\providecommand \href [0]{\begingroup \@sanitize@url \@href}%
\providecommand \@href[1]{\@@startlink{#1}\@@href}%
\providecommand \@@href[1]{\endgroup#1\@@endlink}%
\providecommand \@sanitize@url [0]{\catcode `\\12\catcode `\$12\catcode `\&12\catcode `\#12\catcode `\^12\catcode `\_12\catcode `\%12\relax}%
\providecommand \@@startlink[1]{}%
\providecommand \@@endlink[0]{}%
\providecommand \url  [0]{\begingroup\@sanitize@url \@url }%
\providecommand \@url [1]{\endgroup\@href {#1}{\urlprefix }}%
\providecommand \urlprefix  [0]{URL }%
\providecommand \Eprint [0]{\href }%
\providecommand \doibase [0]{https://doi.org/}%
\providecommand \selectlanguage [0]{\@gobble}%
\providecommand \bibinfo  [0]{\@secondoftwo}%
\providecommand \bibfield  [0]{\@secondoftwo}%
\providecommand \translation [1]{[#1]}%
\providecommand \BibitemOpen [0]{}%
\providecommand \bibitemStop [0]{}%
\providecommand \bibitemNoStop [0]{.\EOS\space}%
\providecommand \EOS [0]{\spacefactor3000\relax}%
\providecommand \BibitemShut  [1]{\csname bibitem#1\endcsname}%
\let\auto@bib@innerbib\@empty
\bibitem [{\citenamefont {Bulmer}\ \emph {et~al.}(2024)\citenamefont {Bulmer}, \citenamefont {Mart{\'\i}nez-Cifuentes}, \citenamefont {Bell},\ and\ \citenamefont {Quesada}}]{lHMT-arxiv24}%
  \BibitemOpen
  \bibfield  {author} {\bibinfo {author} {\bibfnamefont {J.~F.}\ \bibnamefont {Bulmer}}, \bibinfo {author} {\bibfnamefont {J.}~\bibnamefont {Mart{\'\i}nez-Cifuentes}}, \bibinfo {author} {\bibfnamefont {B.~A.}\ \bibnamefont {Bell}},\ and\ \bibinfo {author} {\bibfnamefont {N.}~\bibnamefont {Quesada}},\ }\bibfield  {title} {\bibinfo {title} {Simulating lossy and partially distinguishable quantum optical circuits: theory, algorithms and applications to experiment validation and state preparation},\ }\bibfield  {journal} {\bibinfo  {journal} {arXiv preprint:}\ }\href {https://doi.org/10.48550/arXiv.2412.17742} {10.48550/arXiv.2412.17742} (\bibinfo {year} {2024})\BibitemShut {NoStop}%
\bibitem [{\citenamefont {Bj{\"o}rklund}\ \emph {et~al.}(2019)\citenamefont {Bj{\"o}rklund}, \citenamefont {Gupt},\ and\ \citenamefont {Quesada}}]{Bjorklund-2019-lhaf0}%
  \BibitemOpen
  \bibfield  {author} {\bibinfo {author} {\bibfnamefont {A.}~\bibnamefont {Bj{\"o}rklund}}, \bibinfo {author} {\bibfnamefont {B.}~\bibnamefont {Gupt}},\ and\ \bibinfo {author} {\bibfnamefont {N.}~\bibnamefont {Quesada}},\ }\bibfield  {title} {\bibinfo {title} {A faster hafnian formula for complex matrices and its benchmarking on a supercomputer},\ }\href {https://doi.org/10.1145/3325111} {\bibfield  {journal} {\bibinfo  {journal} {Journal of Experimental Algorithmics}\ }\textbf {\bibinfo {volume} {24}},\ \bibinfo {pages} {1} (\bibinfo {year} {2019})}\BibitemShut {NoStop}%
\bibitem [{\citenamefont {Quesada}\ \emph {et~al.}(2019)\citenamefont {Quesada}, \citenamefont {Helt}, \citenamefont {Izaac}, \citenamefont {Arrazola}, \citenamefont {Shahrokhshahi}, \citenamefont {Myers},\ and\ \citenamefont {Sabapathy}}]{Quesada-PRA2019-lhaf}%
  \BibitemOpen
  \bibfield  {author} {\bibinfo {author} {\bibfnamefont {N.}~\bibnamefont {Quesada}}, \bibinfo {author} {\bibfnamefont {L.~G.}\ \bibnamefont {Helt}}, \bibinfo {author} {\bibfnamefont {J.}~\bibnamefont {Izaac}}, \bibinfo {author} {\bibfnamefont {J.~M.}\ \bibnamefont {Arrazola}}, \bibinfo {author} {\bibfnamefont {R.}~\bibnamefont {Shahrokhshahi}}, \bibinfo {author} {\bibfnamefont {C.~R.}\ \bibnamefont {Myers}},\ and\ \bibinfo {author} {\bibfnamefont {K.~K.}\ \bibnamefont {Sabapathy}},\ }\bibfield  {title} {\bibinfo {title} {{Simulating realistic non-Gaussian state preparation}},\ }\href {https://doi.org/10.1103/PhysRevA.100.022341} {\bibfield  {journal} {\bibinfo  {journal} {Phys. Rev. A}\ }\textbf {\bibinfo {volume} {100}},\ \bibinfo {pages} {022341} (\bibinfo {year} {2019})}\BibitemShut {NoStop}%
\bibitem [{\citenamefont {Kruse}\ \emph {et~al.}(2019)\citenamefont {Kruse}, \citenamefont {Hamilton}, \citenamefont {Sansoni}, \citenamefont {Barkhofen}, \citenamefont {Silberhorn},\ and\ \citenamefont {Jex}}]{Hamilton2019-DetailedGBS}%
  \BibitemOpen
  \bibfield  {author} {\bibinfo {author} {\bibfnamefont {R.}~\bibnamefont {Kruse}}, \bibinfo {author} {\bibfnamefont {C.~S.}\ \bibnamefont {Hamilton}}, \bibinfo {author} {\bibfnamefont {L.}~\bibnamefont {Sansoni}}, \bibinfo {author} {\bibfnamefont {S.}~\bibnamefont {Barkhofen}}, \bibinfo {author} {\bibfnamefont {C.}~\bibnamefont {Silberhorn}},\ and\ \bibinfo {author} {\bibfnamefont {I.}~\bibnamefont {Jex}},\ }\bibfield  {title} {\bibinfo {title} {{Detailed study of Gaussian boson sampling}},\ }\href {https://doi.org/10.1103/PhysRevA.100.032326} {\bibfield  {journal} {\bibinfo  {journal} {Phys.~Rev.~A}\ }\textbf {\bibinfo {volume} {100}},\ \bibinfo {pages} {032326} (\bibinfo {year} {2019})}\BibitemShut {NoStop}%
\bibitem [{\citenamefont {Hamilton}\ \emph {et~al.}(2017)\citenamefont {Hamilton}, \citenamefont {Kruse}, \citenamefont {Sansoni}, \citenamefont {Barkhofen}, \citenamefont {Silberhorn},\ and\ \citenamefont {Jex}}]{Hamilton2017-GBS}%
  \BibitemOpen
  \bibfield  {author} {\bibinfo {author} {\bibfnamefont {C.~S.}\ \bibnamefont {Hamilton}}, \bibinfo {author} {\bibfnamefont {R.}~\bibnamefont {Kruse}}, \bibinfo {author} {\bibfnamefont {L.}~\bibnamefont {Sansoni}}, \bibinfo {author} {\bibfnamefont {S.}~\bibnamefont {Barkhofen}}, \bibinfo {author} {\bibfnamefont {C.}~\bibnamefont {Silberhorn}},\ and\ \bibinfo {author} {\bibfnamefont {I.}~\bibnamefont {Jex}},\ }\bibfield  {title} {\bibinfo {title} {{Gaussian Boson Sampling}},\ }\href {https://doi.org/10.1103/PhysRevLett.119.170501} {\bibfield  {journal} {\bibinfo  {journal} {Phys.~Rev.~Lett.}\ }\textbf {\bibinfo {volume} {119}},\ \bibinfo {pages} {170501} (\bibinfo {year} {2017})}\BibitemShut {NoStop}%
\bibitem [{\citenamefont {Hardy}(2006)}]{Hardy2006FDB}%
  \BibitemOpen
  \bibfield  {author} {\bibinfo {author} {\bibfnamefont {M.}~\bibnamefont {Hardy}},\ }\bibfield  {title} {\bibinfo {title} {Combinatorics of partial derivatives},\ }\bibfield  {journal} {\bibinfo  {journal} {arXiv preprint:}\ }\href {https://doi.org/10.48550/arXiv.math/0601149} {10.48550/arXiv.math/0601149} (\bibinfo {year} {2006})\BibitemShut {NoStop}%
\bibitem [{\citenamefont {Kocharovsky}\ \emph {et~al.}(2022{\natexlab{a}})\citenamefont {Kocharovsky}, \citenamefont {Kocharovsky},\ and\ \citenamefont {Tarasov}}]{LAA2022-HafnianMasterTheorem}%
  \BibitemOpen
  \bibfield  {author} {\bibinfo {author} {\bibfnamefont {V.~V.}\ \bibnamefont {Kocharovsky}}, \bibinfo {author} {\bibfnamefont {{\relax Vl}.~V.}\ \bibnamefont {Kocharovsky}},\ and\ \bibinfo {author} {\bibfnamefont {S.~V.}\ \bibnamefont {Tarasov}},\ }\bibfield  {title} {\bibinfo {title} {{The Hafnian Master Theorem}},\ }\href {https://doi.org/10.1016/j.laa.2022.06.021} {\bibfield  {journal} {\bibinfo  {journal} {Linear Algebra and its Applications}\ }\textbf {\bibinfo {volume} {651}},\ \bibinfo {pages} {144} (\bibinfo {year} {2022}{\natexlab{a}})}\BibitemShut {NoStop}%
\bibitem [{\citenamefont {Isserlis}(1918)}]{Isserlis1918}%
  \BibitemOpen
  \bibfield  {author} {\bibinfo {author} {\bibfnamefont {L.}~\bibnamefont {Isserlis}},\ }\bibfield  {title} {\bibinfo {title} {On a formula for the product-moment coefficient of any order of a normal frequency distribution in any number of variables},\ }\href {https://doi.org/10.2307/2331932} {\bibfield  {journal} {\bibinfo  {journal} {Biometrika}\ }\textbf {\bibinfo {volume} {12}},\ \bibinfo {pages} {134} (\bibinfo {year} {1918})}\BibitemShut {NoStop}%
\bibitem [{\citenamefont {Wick}(1950)}]{Wick1950}%
  \BibitemOpen
  \bibfield  {author} {\bibinfo {author} {\bibfnamefont {G.-C.}\ \bibnamefont {Wick}},\ }\bibfield  {title} {\bibinfo {title} {The evaluation of the collision matrix},\ }\href {https://doi.org/10.1103/PhysRev.80.268} {\bibfield  {journal} {\bibinfo  {journal} {Phys. Rev.}\ }\textbf {\bibinfo {volume} {80}},\ \bibinfo {pages} {268} (\bibinfo {year} {1950})}\BibitemShut {NoStop}%
\bibitem [{\citenamefont {Kocharovsky}\ \emph {et~al.}(2022{\natexlab{b}})\citenamefont {Kocharovsky}, \citenamefont {Kocharovsky},\ and\ \citenamefont {Tarasov}}]{PRA2022-AtomicBS}%
  \BibitemOpen
  \bibfield  {author} {\bibinfo {author} {\bibfnamefont {V.~V.}\ \bibnamefont {Kocharovsky}}, \bibinfo {author} {\bibfnamefont {{\relax Vl}.~V.}\ \bibnamefont {Kocharovsky}},\ and\ \bibinfo {author} {\bibfnamefont {S.~V.}\ \bibnamefont {Tarasov}},\ }\bibfield  {title} {\bibinfo {title} {{Atomic boson sampling in a Bose-Einstein-condensed gas}},\ }\href {https://doi.org/10.1103/PhysRevA.106.063312} {\bibfield  {journal} {\bibinfo  {journal} {Phys.~Rev.~A}\ }\textbf {\bibinfo {volume} {106}},\ \bibinfo {pages} {063312} (\bibinfo {year} {2022}{\natexlab{b}})}\BibitemShut {NoStop}%
\end{thebibliography}%

\end{document}